\documentclass[%
reprint,
showpacs,
 amsmath,amssymb,
 aps,
 pre,
]{revtex4-1}

\usepackage{graphicx}
\usepackage[percent]{overpic}
\usepackage{dcolumn}
\usepackage{bm}

\usepackage{hyperref}
\usepackage{color} 
\definecolor{rltred}{rgb}{0.75,0,0}
\definecolor{rltgreen}{rgb}{0,0.6,0}
\definecolor{rltblue}{rgb}{0.3,0.3,1}

\hypersetup{colorlinks,%
    hypertexnames=true,%
    linkcolor=rltred,%
    citecolor=rltgreen,%
    linktocpage=true}
    
\begin{document}
\title{Semiclassical wavefunctions for open quantum billiards}

\author{Fabian Lackner}
\email{fabian.lackner@tuwien.ac.at}
\affiliation{Institute for Theoretical Physics, Vienna University of Technology,
Wiedner Hauptstra\ss e 8-10/136, 1040 Vienna, Austria, EU} 

\author{Iva B\v rezinov\'a}
\email{iva.brezinova@tuwien.ac.at}
\affiliation{Institute for Theoretical Physics, Vienna University of Technology,
Wiedner Hauptstra\ss e 8-10/136, 1040 Vienna, Austria, EU} 

\author{Florian Libisch}
\affiliation{Department of Mechanical and Aerospace Engineering, Princeton University, New Jersey,USA} 

\author{Joachim Burgd\"orfer}
\affiliation{Institute for Theoretical Physics, Vienna University of Technology,
Wiedner Hauptstra\ss e 8-10/136, 1040 Vienna, Austria, EU}  

\date{\today}

\begin{abstract}
We present a semiclassical approximation to the scattering wavefunction $\Psi(\mathbf{r},k)$ for an open quantum billiard which is based on the reconstruction of the Feynman path integral. We demonstrate its remarkable numerical accuracy for the open rectangular billiard and show that the convergence of the semiclassical wavefunction to the full quantum state is controlled by the path length or equivalently the dwell time. Possible applications include leaky billiards and systems with decoherence present.
\end{abstract}
\maketitle

\section{\label{sec:Int}Introduction}
In his doctoral thesis R. P. Feynman, extending earlier work by Dirac \cite{Dir33}, developed a novel formulation of quantum mechanics \cite{Fey48}.
Unlike Schr\"odinger's formulation in terms of the solution of a partial differential equation Feynman based his description on the intuitive picture of paths connecting two points in space. Each path carries an amplitude and a phase that is given by its classical action. Incorporating the principle of superposition, the propagator, i.e.~the probability amplitude to move from one point in space to another, is given by the sum (i.e.~integral) over all paths connecting these points. While this path integral formulation is equivalent to the standard Schr\"odinger theory, its implementation as an operational algorithm to solve quantum problems is complicated due the mathematical difficulties associated with the path integration. One of its advantages is, however, the conceptual insight it can provide. Most importantly, modern semiclassical theory invokes the convergence of the path manifold as contained in Feynman's path integral towards a discrete subset of classical paths of extremal action as $\hbar\rightarrow 0$. 

The semiclassical approximation applicable at the border between quantum and classical mechanics follows from the path integral formalism in the limit that the variation of the classical action is large compared to $\hbar$ for small path variations. Mesoscopic systems with linear dimension $D$ large compared to the de Broglie wavelength $\lambda_{dB}$, $\lambda_{dB}\ll D$, represent prototypical cases for which semiclassical approximations are frequently invoked since in many cases ab-initio quantum calculations become unfeasible. Moreover, the description in terms of paths can provide detailed physical insights into spectral and transport properties. For example, dephasing and decohering interactions are associated with a characteristic mean free pathlength $\ell_{\rm MFP}$, thereby limiting phase coherent transport to short paths, $\ell<\ell_{\rm MFP}$. A well known example is the open quantum billiard in the ballistic regime which has been extensively studied in the last several years both experimentally and theoretically (see, e.g., Ref.~\onlinecite{Bee97,Dat95,AkkMon06} and references therein). The semiclassical approximation has contributed to the understanding of phase coherent transport effects such as conductance fluctuations and weal localization  \cite{IshBur95,BloZoz00,BraBha03,BluSmi90,BarJalSto93,BloZoz01,WirTanBur97,WirTanBur99,WirStaRotBur03,StaRotBurWir05,IvaWirRotBur10,BreStaWirRotBur08,RahBro06,RahBro05,JacWhi06,RicSie02,BraHeuMulHaa06,HeuMulBraHaa06,BroRah06,PicJal99,Bog00}. Most semiclassical approximations to date have focused on either the spectral density $\rho(E)$ \cite{Sto06} or transport coefficients determined for ballistic transport by $S$ matrix elements \cite{BarJalSto93}. Semiclassical calculations of the wavefunction $\Psi(\mathbf{r},k)$ itself which test the quantum to classical transition locally on the finest scale have remained a challenge. Few pioneering studies have been performed: scars in closed billiards could be reproduced by semiclassical calculations of the energy-averaged probability density $\langle |\Psi(\mathbf{r},k)|^2\rangle$ \cite{Bog87}.  For open chaotic billiards, statistical properties of the wavefunctions like nodal point distributions have been found in good agreement with random wave models \cite{HoeKuhSto09}. In the regime of high incident energies with a large number of open modes, quantum calculations \cite{IshKea04} for the open chaotic stadium billiard have shown that the wavefunction closely mirrors the path bundles \cite{WirTanBur97} of short classical scattering trajectories. A simple semiclassical approximation to the wavefunction yields good quantiative agreement with the quantum wavefunction. \\
We present in the following an accurate semiclassical determination of the fine-scale wavefunction $\Psi(\mathbf{r},k)$ for an open ballistic billiard. We construct the wavefunction in terms of a sum over paths connecting the entrance lead with an arbitrary point $\mathbf{r}$ in the interior of the billiard closely following the Feynman path integral prescription. We employ the pseudopath semiclassical approximation (PSCA) \cite{WirStaRotBur03,StaRotBurWir05,BreStaWirRotBur08,IvaWirRotBur10} to include both classical and diffractive, i.e.~non-classical paths into the path sum. We aim at a quantitative agreement with quantum wavefunctions for the low-energy regime with only few open modes in the leads (quantum wires) and a semiclassical description that pertains to the interior of the billiard. We gauge the accuracy of the wavefunction by comparison with full quantum wavefunctions. For technical reasons, we focus on the rectangular billiard for which the enumeration and summation of paths is still feasible since it is a prototypical example of a integrable system. We show that the convergence towards the quantum wavefunction is controlled by the dwell time, or equivalently, by the mean pathlength of the scattering state at given energy.\\
The outline of the paper is as follows: In Sec.~\ref{sec:sem} we briefly review the PSCA of the constant energy propagator and present its extension to calculations of wavefunctions of open quantum billiards. Concurrent calculations of the full quantum scattering state as well as its truncated form in which Fourier components associated with long path length are filtered out are discussed in Sec.~\ref{sec:qm}. In Sec.~\ref{sec:num} we present a quantitative comparison between semiclassical and quantum wavefunctions. We relate the convergence of the semiclassical wavefunction towards the exact quantum wavefunction to the dwell time, or, equivalently, to the Eisenbud-Wigner-Smith (EWS) delay time of the scattered wave inside the ballistic cavity. In the outlook (Sec.~\ref{sec:sum}) we briefly discuss future applications to leaky billiards and decoherence processes where long (coherent) paths or delay times are effectively suppressed and the PSCA offers a simple route to construct wavefunctions when short paths dominate the dynamics.
%
\section{\label{sec:sem}Semiclassical theory for scattering states}
\subsection{\label{sec:billiard}The rectangular ballistic billiard}

\begin{figure*}[t]
\includegraphics[width=\textwidth]{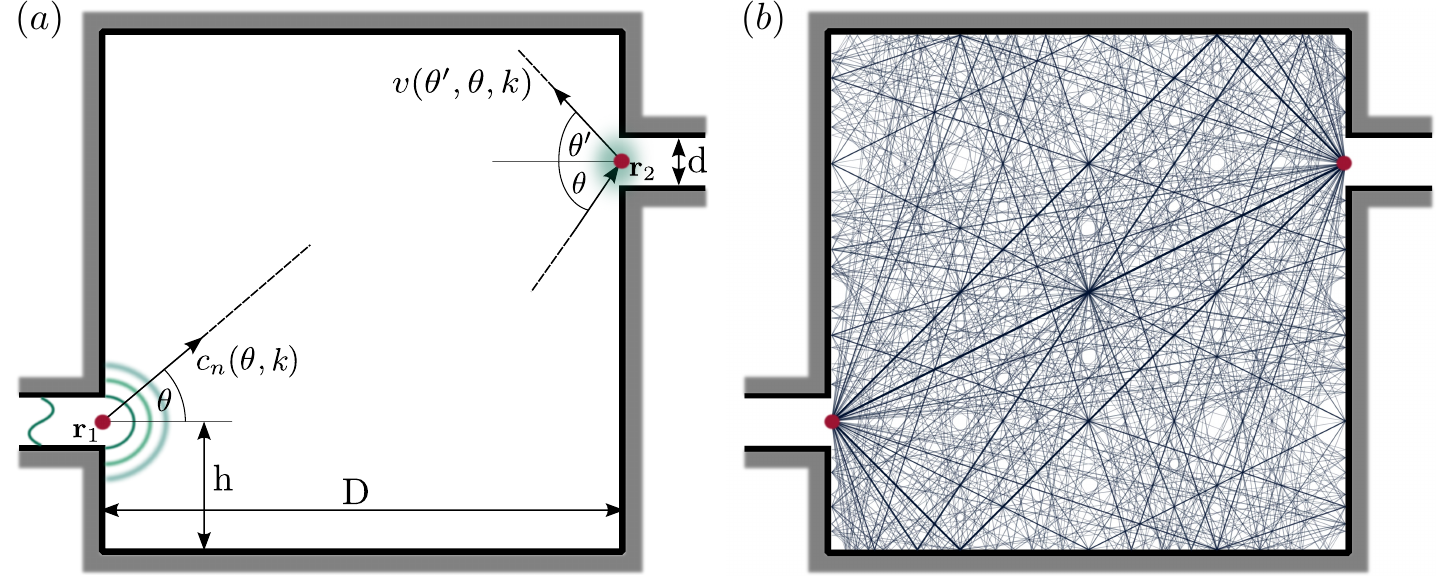}
\caption{(Color online) (a) Geometry of the rectangular (square) billiard with side length $D=1$ and equal lead widths $d=1/16$. The placement of the leads is point symmetric with offset $h=0.25$. The diffractive couplings from the leads (quantum point contacts) to the cavity, $c_n(\theta,k)$, and for backscattering into the cavity $v(\theta^{\prime},\theta,k)$ are sketched. The quantum point contacts (red dots) placed in the center of the leads are located at the coordinates $(x_1=0,y_1=0)$ and $(x_2=D,y_2=D-2h)$. (b) The set of classical paths that connect the two quantum point contacts up to the maximal length $L=10$. The color intensity of the paths is proportional to their deflection factor defined in Eq.~\ref{eq:Dp}.}
\label{fig:geo}
\end{figure*}

We discuss in the following the PSCA for wavefunctions with the help of the specific example of a squared (rectangular) open billiard [Fig.~\ref{fig:geo} (a)]. The lead width $d=1/16$ is small compared to the linear dimension $D=1$ in reduced units. Consequently, the short-wave limit $\lambda_{dB}\ll D$ is reached for the internal dynamics of the cavity while the motion inside the leads (or quantum point contacts) is still in the quantum regime for low transverse mode numbers $n$ with wavenumber $k_n=n\pi/d$. All wavenumbers are given in the following in units of $\pi/d$. The PSCA is designed to describe the semiclassical dynamics for billiards which are coupled to quantum wires. The asymptotic scattering boundary condition is defined by the incoming wave
\begin{eqnarray}\label{eq:incident_current}
\Psi_{n,k_x}(\mathbf{r})=\frac{1}{\sqrt{k_n}}\chi_n(y)e^{ik_nx},
\end{eqnarray}
where $(x,y)$ are the local coordinates along and perpendicular to the entrance lead and $\chi_n(y)=\sqrt{\frac{2}{d}}\sin{[\frac{n\pi}{d} (y-\frac{d}{2})]}$ is the transverse wavefunction of mode $n$. $\Psi_{n,k_x}(\mathbf{r})$ is flux normalized. The center of our coordinate system is the quantum point contact at the entrance lead denoted by $\mathbf{r_1}=(x=0,y=0)$ (see Fig.~\ref{fig:geo}). The energy of the scattering state is given by $E=k^2/2$ with $k=\sqrt{k_x^2+(n\pi/d)^2}$. The potential in the interior of the billiard cavity vanishes, $V=0$, and is infinitely high at the walls. Dephasing and decohering interactions with the environment are neglected.\\
The classical dynamics of transport from the entrance to the exit lead (transmission) or back to the entrance (reflection) is given by all classical paths $p$ connecting the quantum point contacts (located at the center of the lead junctions) with each other. The action of a classical path $p$ with path length $L_p$ is given by $S_p=kL_p$. The classical phase space is structured in paths (or path bundles \cite{WirTanBur97} when the finite size of the lead openings is taken into account) whose weight (area in phase space) is given by the deflection factor
\begin{align}\label{eq:Dp}
D_p(k)=\begin{vmatrix}
\frac{\partial ^2 S}{\partial \mathbf{r^{\prime}} \partial \mathbf{r}},\frac{1}{k}\frac{\partial ^2 S}{\partial \mathbf{r} \partial k} \\ \frac{1}{k}\frac{\partial ^2 S}{\partial \mathbf{r^{\prime}} \partial k},\frac{1}{k^2}\frac{\partial ^2 S}{\partial k^2}
\end{vmatrix} .
\end{align}
For the rectangular billiard the deflection factor is exactly $D_p=1/(kL_p)$. In Fig.~\ref{fig:geo} (b) we depict several classical paths connecting the entrance and exit point contact. Each path represents a path bundle and its color intensity is proportional to its weight, i.e.~the deflection factor. For the rectangular billiard $D_p$ is strictly positive and free of singularities. Consequently, the Maslov indev $\mu_p$ entering semiclassics is given by 
\begin{equation}
\mu_p=2N_p
\end{equation}
where $N_p$ is the number of reflections from the hard wall along path $p$. Note that one could alternatively incorporate von Neumann boundary conditions by setting the Maslov index equal to zero.

\subsection{\label{sec:PSCA}The PSCA for scattering states}
Starting point for the development of a semiclassical approximation to the wavefunction of the scattering state in the interior of the billiard is its expression in terms of the Green's function $G(\mathbf{r^{\prime}},\mathbf{r},k)$,
\begin{equation}\label{eq:psi}
\Psi_n(\mathbf{r^{\prime}},k)=-i\sqrt{k_n}\int_{-d/2}^{d/2}G(\mathbf{r^{\prime}},x=0,y)\chi_n(y)dy,
\end{equation}
for an incoming particle with wavenumber $k=\sqrt{2E}$ in mode $n$. The scattering boundary condition (Eq.~\ref{eq:incident_current}) at the lead entrance $(x=0,y)$ acts as a source and the Green's function or constant-energy Feynman propagator connects the entrance point with the observation point via all (non-classical) paths. Eq.~\ref{eq:psi} contains an integral over all entrance points with transverse coordinate $y$.\\
We note that the Fisher-Lee equation \cite{FisLee81} expressing the $S$ matrix elements in terms of $G$ is a special case of Eq.~\ref{eq:psi}. For example the transmission amplitude $t_{mn}$ is given by 
\begin{align} 
\label{eq:greent}
t_{mn}(k)&=-i\sqrt{k_m k_n}\\ \nonumber
&\times\int\int\chi_{m}^*(y^{\prime}) G(x'=D,y',x=0,y;k) \chi_{n}(y) \text{d}y^{\prime} \text{d}y,
\end{align}
where the integral over $y'$ extends over all coordinates at the exit lead weighted with the outgoing mode amplitude $\chi_m(y')$ in the exit channel. The transmission (reflection) matrix is related to the $S$ matrix as $t_{mn}=S^{2,1}_{mn}$ ($r_{mn}=S^{1,1}_{mn}$) and the conductance $g$ of the quantum billiard is determined by the Landauer formula \cite{Lan87,Lan57}:
\begin{equation}\label{eq:Landauer}
g(k)=\frac{2e^2}{h}T(k)=\frac{2e^2}{h}\sum_{n=1}^{N}\sum_{m=1}^{N}\vert t_{mn}(k)\vert ^2,
\end{equation}  
where $N$ is the number of open modes in the leads (in our case the leads have equal width).\\
The semiclassical approximation to Eq.~\ref{eq:psi} and Eq.~\ref{eq:greent} involves two steps: First, the evaluation of the integrals over the entrance/exit leads by either a stationary phase approximation (SPA) as done in the conventional semiclassical approximation or as a diffractive integral \cite{WirTanBur97} as implemented in the PSCA. We employ for the latter a combination of the geometric theory of diffraction \cite{Kel62} (GTD) and the uniform theory of diffraction \cite{KouPat74,SiePavSch96} (UTD), called GTD-UTD, which has been previously successfully applied for the scattering matrix of the circular billiard \cite{IvaWirRotBur10}. The GTD-UTD takes into account the multiple scattering between the edges of a given lead (for details see \cite{IvaWirRotBur10}). Second, the quantum propagator is replaced within the PSCA by a semiclassical propagator that contains in addition to classical paths also pseudopaths, i.e.~sequences of classical paths joined by diffractive (back) scattering at the lead openings. The reasoning underlying this augmented path manifold is that near the sharp edges of the leads or, more generally, at the interface between the quantum point contacts and the cavity (Fig.~\ref{fig:geo}), the semiclassical regime of sufficiently small $\lambda_{dB}$ cannot be reached and, therefore, non-classical path contributions, determined to leading order in $\hbar$ by diffractive integrals, must be included from the outset. These pseudopaths are intuitively simple realizations of contributions to the Feynman path integral that are non-classical in origin, yet can be systematically included in the approximation. Accordingly, each pseudopath $p$ of order $\eta(p)$ consists of a sequence of $\eta+1$ classical path segments $p_i$ joined by $\eta$ diffractive scatterings at one of the lead openings (or point contacts) with diffractive amplitude $v(\theta^{p_i},\theta^{p_i-1},k)$, where $\theta^{p_i-1}$ ($\theta^{p_i}$) are the incoming (outgoing) scattering angle with which the trajectory $p_{i-1}$ ($p_i$) approaches (leaves) the point contact. Explicit analytic expressions of the coupling coefficient $v(\theta^{p_i},\theta^{p_i-1},k)$ are given in appendix~\ref{sec:app}. The PSCA to the propagation along the pseudopath $p$ between the starting point $\mathbf{r}$ and the end point $\mathbf{r'}$ reads
\begin{align}\label{eq:GPSCA_part}
G^{\rm PSCA}_{p}(\mathbf{r'},\mathbf{r},k)=\left[\prod_{i=1}^\eta G_{p_i}^{\rm SCA}(k)v(\theta^{p_i},\theta^{p_{i-1}},k)\right]G_{p_0}^{\rm SCA}(k),
\end{align}
where the amplitude for each classical path segment $p_i$ connecting two quantum point contacts is given by the standard SCA expression for the rectangular billiard
\begin{align}
\label{eq:semig}
G^{\rm SCA}_{p}(k)
&=\dfrac{2 \pi}{(2 \pi i)^{3/2}}\sqrt{\vert D_p(k) \vert }\times \\ \nonumber
&\exp\left[i S_p(k)-i\dfrac{\pi}{2}\mu_p \right] \\ \nonumber
&=\frac{2\pi}{(2\pi i)^{3/2}}\frac{1}{\sqrt{kL_{p}}}\exp\left[ikL_{p}-iN_{p}\pi\right].
\end{align}

The complete propagator $G_{\Lambda}^{\rm PSCA}(\mathbf{r}',\mathbf{r},k)$ to order $\Lambda$ is the sum over all contributions from pseudopaths connecting $\mathbf{r^{\prime}}$ and $\mathbf{r}$ with $\eta(p)\leq\Lambda$:
\begin{align}\label{eq:GPSCA}
G_{\Lambda}^{\rm PSCA}(\mathbf{r^{\prime}},\mathbf{r},k)=\sum_{p:\eta(p) \leq \Lambda}G^{\rm PSCA}_{p}(\mathbf r',\mathbf r,k).
\end{align}

Note that in Eq.~\ref{eq:GPSCA} $\mathbf{r^{\prime}}$ and $\mathbf{r}$ can be arbitrary points inside the billiard. In the application to the wavefunction Eq.~\ref{eq:psi} we will set $\mathbf{r}$ to be the entrance point contact $\mathbf{r}_1$ and $\mathbf{r'}$ to be an arbitrary internal point inside the billiard while in Eq.~\ref{eq:greent} $\mathbf{r'}$ is one of the lead-billiard junctions when transmission or reflection is determined. 
The order $\Lambda$ controls the degree to which diffractive contributions are included. Since the sum in Eq.~\ref{eq:GPSCA} extends over infinitely many contributions its numerical evaluation requires in practice the limitation of path lengths by a cut-off length $L_{\rm max}$. For integrable billiards such as the rectangular billiard the number of trajectories below a maximum length $M^{\rm SCA}(L_{\rm max})$ increases quadratically, $M^{\rm SCA}(L_{\rm max})\propto L_{\rm max}^2$. However, the number of pseudopaths resulting from joining classical paths by a sequence of diffractive couplings eventually proliferates exponentially, $M^{\rm PSCA}(L_{\rm max})\propto \exp{(L_{\rm max}/L_0)}$. Therefore, sums over pseudopaths in numerical implementations can only by executed up to modest length $L_{\rm max}$. The convergence depends on the parameter pair $(\Lambda,L_{\rm max})$. Note that $G_{\Lambda=0}^{\rm PSCA}$ (Eq.~\ref{eq:GPSCA}) is not equivalent to $G^{\rm SCA}$ (Eq.~\ref{eq:semig}). In the SCA a path that hits the exit leaves the cavity. Within the PSCA the diffractive scattering at the exit gives rise to a plane wave and, to first order, a circular diffractive wave. In zeroth order the circular wave vanishes but the plane wave leads to geometrically reflected paths which are not included in the SCA.\\ 
The propagator within the PSCA, $G^{\rm PSCA}$, can be used to construct the semiclassical wavefunction via Eq.~\ref{eq:psi}. However, the integral over the entrance lead-billiard junction weighted with the transverse mode wavefunction $\chi_n$ would require the calculation of $G^{\rm PSCA}(\mathbf{r}^\prime,x=0,y)$ for all points $-d/2<y<d/2$. For low mode numbers $n$ we can approximate the integral by a diffraction approximation that replaces the lead junction by a quantum point contact which acts as a point scatterer located at $\mathbf{r}_1$. 

We note that this description is appropriate for low-energy scattering with $\lambda_{\text{dB}} \approx d$ while in the high energy limit $\lambda_{\text{dB}} \ll d$ entire path bundles emanating from the finite-size lead opening rather than resolved paths conecting the point contacts should be included in the semiclassical approximation \cite{IshKea04}.  In the following we will focus on paths emitted from this point contact. The amplitude for a incoming particle in mode $n$ to leave the point contact at the entrance lead with launching angle $\theta$ relative to the lead axis (see Fig.~\ref{fig:geo}) and wavenumber $k$ is denoted by $c_n(\theta,k)$ (see appendix~\ref{sec:app} for its analytic form). Analogously $c^*_m(\theta,k)$ represents the amplitude for a trajectory incident on the junction under the angle $\theta$ to exit in mode $m$.\\
Inserting Eq.~\ref{eq:GPSCA} and $c_n(\theta,k)$ into Eq.~\ref{eq:psi} yields the PSCA for the scattering wavefunction inside the billiard subject to the boundary condition (Eq.~\ref{eq:incident_current}) of incident current in mode $n$
\begin{align}\label{eq:psi_psca}
\Psi_{n,\Lambda}^{\rm PSCA}(\mathbf{r}',k)=&-i\sqrt{k_n} \\ \nonumber
&\times\sum_{p:\eta(p)\leq\Lambda}G_p^{\rm PSCA}(\mathbf r',\mathbf r_1,k)c_n(\theta_p^e,k),
\end{align}
where $\theta_p^e$ is the entrance angle of the trajectory $p$. Analogously, the transmission amplitude follows as
\begin{align}\label{eq:trans}
t_{mn}(k)=&-i\sqrt{k_m k_n} \\ \nonumber
&\times\sum_{p:\eta(p)\leq\Lambda}c_m(\theta_p^{f},k) G_{p}^{\rm PSCA}(\mathbf r_2,\mathbf r_1,k) c_n(\theta_p^{e},k).
\end{align}
We will use in the following Eq.~\ref{eq:trans} as a complementary test for the accuracy of the PSCA to the wavefunction in Eq.~ \ref{eq:psi_psca}.
\section{Quantum calculations} \label{sec:qm}
Before presenting results of the PSCA, we briefly review the method employed for solving the underlying quantum problem. Our quantum calculations are based on the modular recursive Green's function method (MRGM)\cite{RotWeiLib06,RotTanWir00,LibRotBur12}. In the MRGM the two-dimensional (2D) Sch\"odinger equation is solved numerically on a tight-binding grid, which leads to a non-quadratic (cosine) dispersion relation, the main source of limitations within tight-binding approximations of the 2D Schr\"odinger equation. In the PSCA the dispersion relation has the correct form of $E=k^2/2$. Thus, the comparison between the PSCA and QM requires small grid spacings within the MRGM. We have used $30$ grid points per half-wave length. A coarser discretization below this value leads to a visible shift in $k$ of the QM transport results with respect to the PSCA due to the non-quadratic dispersion relation.\\
Even though the concept of paths and path lengths does not explicitly enter the quantum description, the scattering wavefunction $\Psi(\mathbf r,k)$ can be Fourier analyzed in terms of its length component $L$,
\begin{align}\label{eq:ft_psi}
\tilde{\Psi}(\mathbf r,L)=\int \text{d}k\Psi(\mathbf r,k)e^{-ikL}.
\end{align}
As we will show the Fourier conjugate variable to $k$ is closely related to the physical variable length of the (semi)classical dynamics. For the numerical evaluation of Eq.~\ref{eq:ft_psi}, we perform a windowed Fourier transform in interval $[k_{\rm min},k_{\rm max}]$.\\
Eq.~\ref{eq:ft_psi} is the generalization of the path length spectroscopy \cite{BloSchZoz02,IshBur95,WirTanBur97,StaRotBurWir05} of $S$ matrix elements
\begin{equation}\label{eq:Smn}
\tilde{S}_{mn}(L)=\int_{k_{\text min}}^{k_{\text max}} \text{d}k \; S_{mn}(k) e^{-ikL}.
\end{equation}
$\tilde{S}_{mn}(L)$ is the probability amplitude for a quantum path of length $L$ to scatter from mode $n$ to $m$. The path-length spectrum of open quantum systems decays for increasing length $L$. However, contributions from very long paths may become important near resonances of long-lived quasi-bound states. We will present examples of path length distributions entering $\Psi$ below.\\
Assuming for the moment that the Fourier component $\tilde\Psi(\mathbf r,L)$ can be, indeed, identified with the semiclassical path length, it is now instructive to construct truncated quantum wavefunctions that retain only Fourier components with $L$ less than the maximum path length $L_{\rm max}$ included in the PSCA. To this end, we truncated the inverse Fourier transform at $L=L_{\rm max}$
\begin{align}\label{eq:trunc_psi}
\Psi^{\rm T}(\mathbf r, k)=\int_0^{L_{\rm max}}\text{d}L\; \tilde\Psi(\mathbf r,L)e^{ikL}.
\end{align}
Analogous truncation of Fourier spectra are performed for $S$ matrix elements
\begin{align}\label{eq:trunc_Smn}
S_{mn}^{\rm T}(k)=\int_0^{L_{\rm max}}\text{d}L\; \tilde S_{mn}(L)e^{ikL}.
\end{align}
Normalizations have been omitted for simplicity. The comparison between $\Psi^{\rm T}$ and $\Psi^{\rm PSCA}$ allows to directly and quantitatively compare the semiclassical wavefunction with the quantum wavefunction that contains all Feynman paths up to the same length $L_{\rm max}$. Conversely, comparison between $\Psi^{\rm T}$ and the full quantum wavefunction $\Psi$ allows to assess the influence of long paths $L>L_{\rm max}$ and, therefore, the truncation error involved in semiclassical path sums (see Sec.~\ref{subsec:wf}).\\
The numerical evaluation of the Fourier transform for finite discretized intervals in $k$ gives rise to a maximal resolvable length $\Delta L=2 \pi /\delta k$ where $\delta k$ is the grid spacing in  the $k$ domain. In order for contributions with $L>L_{\rm max}$ not to enter $\Psi^{\rm T}$, the amplitude at the maximum resolvable length $\Delta L$, $\tilde\Psi(\mathbf{r},\Delta L)$, must already be strongly suppressed. Otherwise, the Fourier spectrum is back-folded such that contributions with $L\gtrsim \Delta L\gg L_{\rm max}$ appear near the origin ($L\approx0$) and cannot be cleanly cut off by Eq.~\ref{eq:trunc_psi}. To avoid such back-folding we choose a large $\Delta L$. However, we will see in Sec.~\ref{subsec:wf} that a complete truncation cannot be established in the vicinity of sharp resonances. This truncation is known as the sinc filter in signal processing. It leads to a smoothing of sharp peaks on the $k$ scale and causes a violation of unitarity since high Fourier components corresponding to contributions from long paths are missing. We note that the energy average of the local density of states in the semiclassical description of scars \cite{Bog87} corresponds to an implicit truncation scheme.\\
The relative importance of long paths can be quantified by the expectation value of the Eisenbud-Wigner-Smith (EWS) time delay operator for incoming mode $n$
\begin{align}
\langle Q\rangle&=\langle -iS^\dagger\frac{\partial}{\partial E}S\rangle \nonumber \\
&= -i \sum_m t_{nm}^\dagger\frac{\partial}{\partial E} t_{mn}+ r_{nm}^\dagger
\frac{\partial}{\partial E}r_{mn}.
\end{align}
For billiards with zero potential in the interior, the time delay $\tau_{\rm EWS}$ can be directly converted into a path length 
\begin{equation}\label{eq:meanlength}
\ell_{\rm EWS}=k \tau_{\rm EWS}.
\end{equation}
By comparing $\ell_{\rm EWS}(k)$ at at given wavenumber (or energy) with $L_{\rm max}$ we can provide an independent estimate for the expected proximity of $\Psi^{\rm T}$ to $\Psi$ and, in turn, the convergence of $\Psi^{\rm PSCA}$ to the full scattering state. \\
An alternative measure for the time the particle spends inside the cavity is the dwell time
\begin{align}\label{eq:tauD}
\tau_{\rm D}=\int\limits_{\rm cavity} |\Psi(\mathbf{r},k)|^2 \text{d}x\;\text{d}y.
\end{align}
The difference between $\tau_{\rm EWS}$ and $\tau_{\rm D}$ is referred to as the interference delay that the wave packet experiences before entering the cavity due to interference with parts of itself that have already been reflected \cite{Win03}. This self-interference delay becomes important when the de Broglie wavelength of the particle is comparable to the linear dimension of the billiard. In the semiclassical regime, however, we find $\tau_{\rm D}=\tau_{\rm EWS}$ to a very good degree of approximation.\\

\section{Numerical results}\label{sec:num}
\subsection{Scattering matrix}\label{subsec:scat_mat}
%
\begin{figure*}[t]
\includegraphics[width=\textwidth]{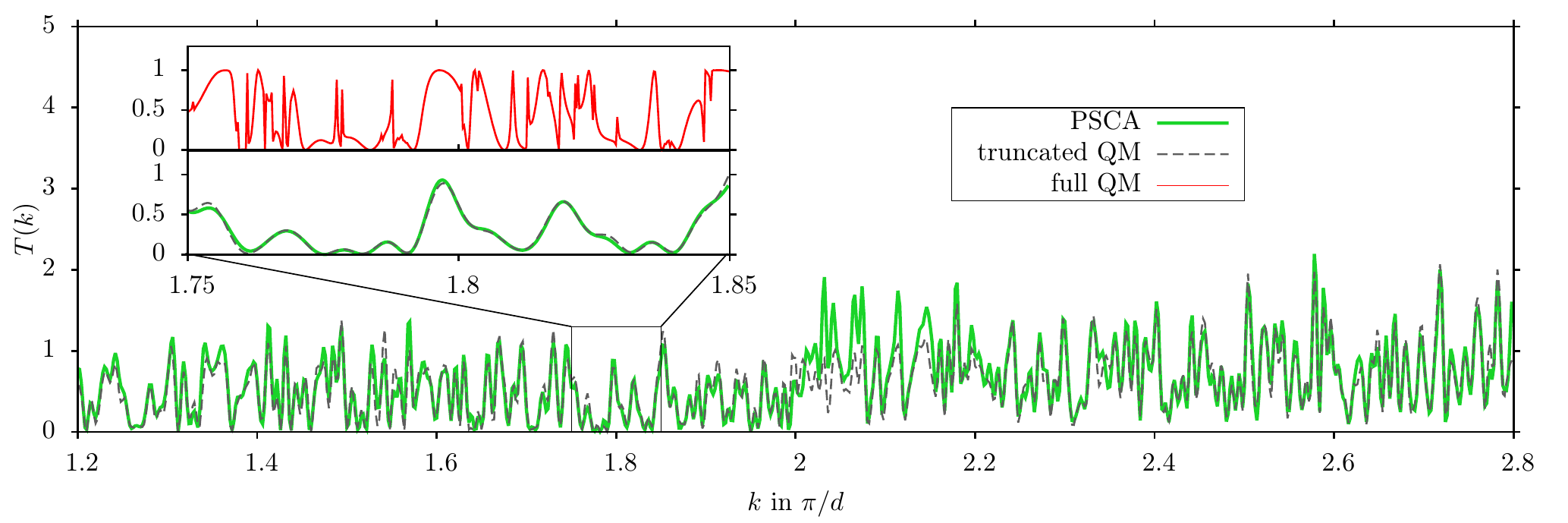}
\caption{(Color online) Comparison of the total transmission $T(k)$ as a function of $k$ within the PSCA (green line), truncated QM (black dashed line) and full QM (in the inset, upper panel red line). The PSCA is calculated with a maximum path length of $L_{\rm max}=17.5$ and maximum order of diffractive scattering $\Lambda=6$. The inset shows a magnification of the region $k \in [1.75,1.85]$.}
\label{fig:smat}
\end{figure*}
To set the stage, we first present typical results for $S$ matrix elements (Eq.~\ref{eq:greent} and Eq.~\ref{eq:trans}). The total transmission $T(k)$ (Eq.~\ref{eq:Landauer}) as a function of the incident energy or, equivalently, $k$ (Fig.~\ref{fig:smat}) displays excellent agreement between the PSCA and the scattering matrix $S^{\rm T}$ truncated at the same path length $L_{\rm max}=17.5$ as the PSCA. A few exceptions are worth mentioning. They appear, e.g, in the vicinity of the channel opening $k\lesssim2\pi/d$. Just above the channel opening the emission angle $\theta$ is close to $\pi/2$. For grazing incidence the diffraction amplitude $c_n(\theta,k)$ and scattering amplitude $v(\theta',\theta,k)$ in GTD-UTD is less accurate, most likely causing this discrepancy. While overall the agreement between the PSCA and the truncated quantum $S$ matrix is remarkable, pronounced differences appear to the full quantum $S$ matrix as highlighted in Fig.~\ref{fig:smat} for the magnified interval $1.75\leq k\leq1.85$. Pronounced sharp structures are missing in the PSCA indicating that the latter are due to long paths, i.e.~represent long-lived resonances. In principle, the PSCA could account for those if $L_{\rm max}$ could be extended. In practice, however, the exponential proliferation of contributing pseudopaths prevents to perform complete path sums. The comparison suggests that for those $k$ values where sharp resonances appear, the PSCA wavefunction will be significantly different from the quantum scattering state $\Psi$ while in spectral regions where $T(k)$ is smooth, the PSCA should become accurate.
\subsection{Wavefunctions}\label{subsec:wf}
%
\begin{figure}
\includegraphics[width=0.39\textwidth]{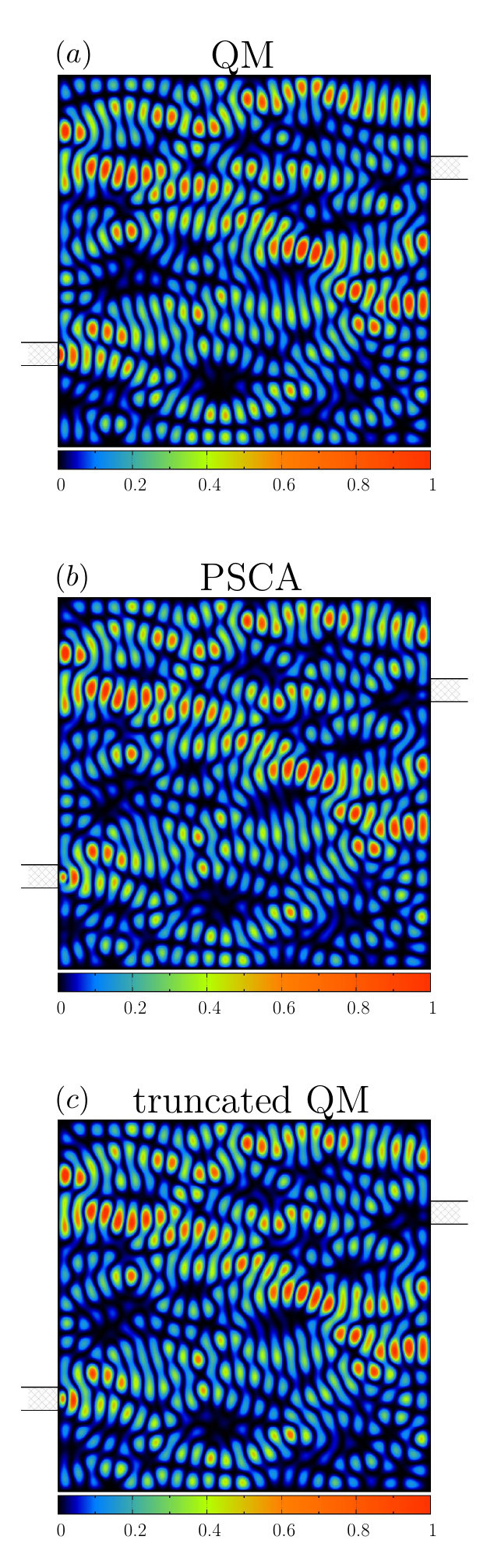}
\caption{(Color online) Comparison of $\Psi^{\rm PSCA}$, the truncated wavefunction $\Psi^{\rm T}$, and the full wavefunction $\Psi$ for a non-resonant scattering state in incident mode $n=1$ with $k=1.7835$ and $\ell_{\rm EWS}=10.5$. (a) Full quantum wavefunction $\Psi$, (b) truncated quantum wavefunction with Fourier components $L\le L_{\text{max}}=17.5$, (c) PSCA with $L_{\text{max}}=17.5$ with maximum diffractive order $\Lambda=6$.}
\label{fig:wavenonreso}
\end{figure}
\begin{figure}
\includegraphics[width=0.39\textwidth]{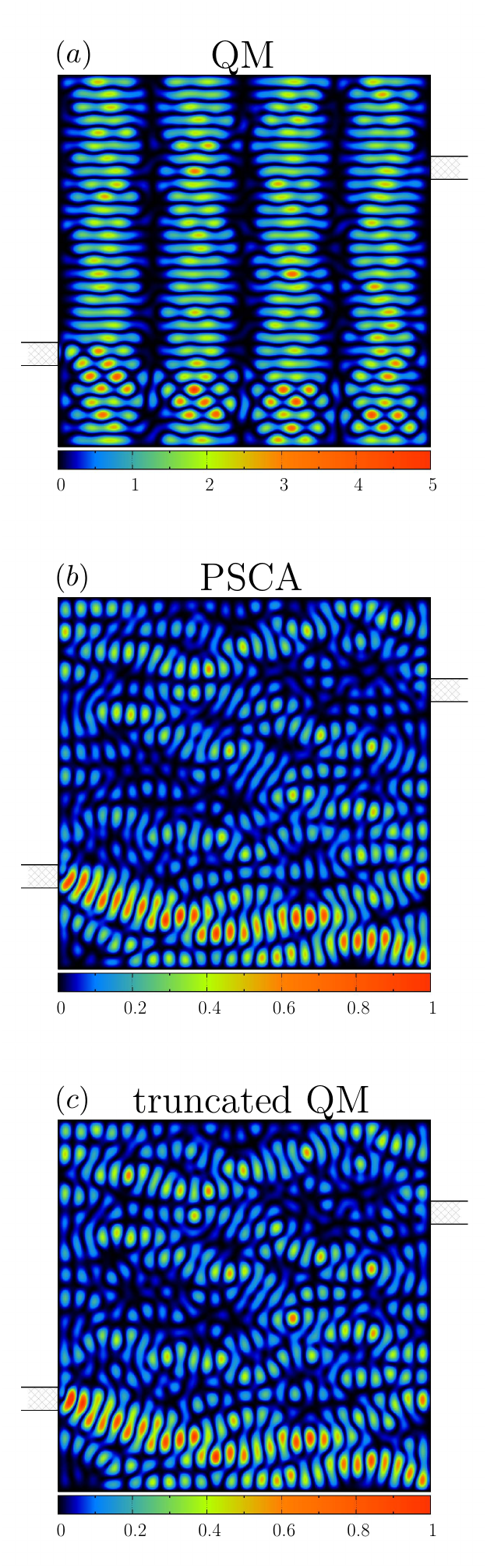}
\caption{(Color online) As Fig.~\ref{fig:wavenonreso} but for a resonant scattering state in incident mode $n=1$ with $k=1.8295$ and $\ell_{\rm EWS}=47$.}
\label{fig:wavereso}
\end{figure}
\begin{figure}[b]
\includegraphics[width=0.5\textwidth]{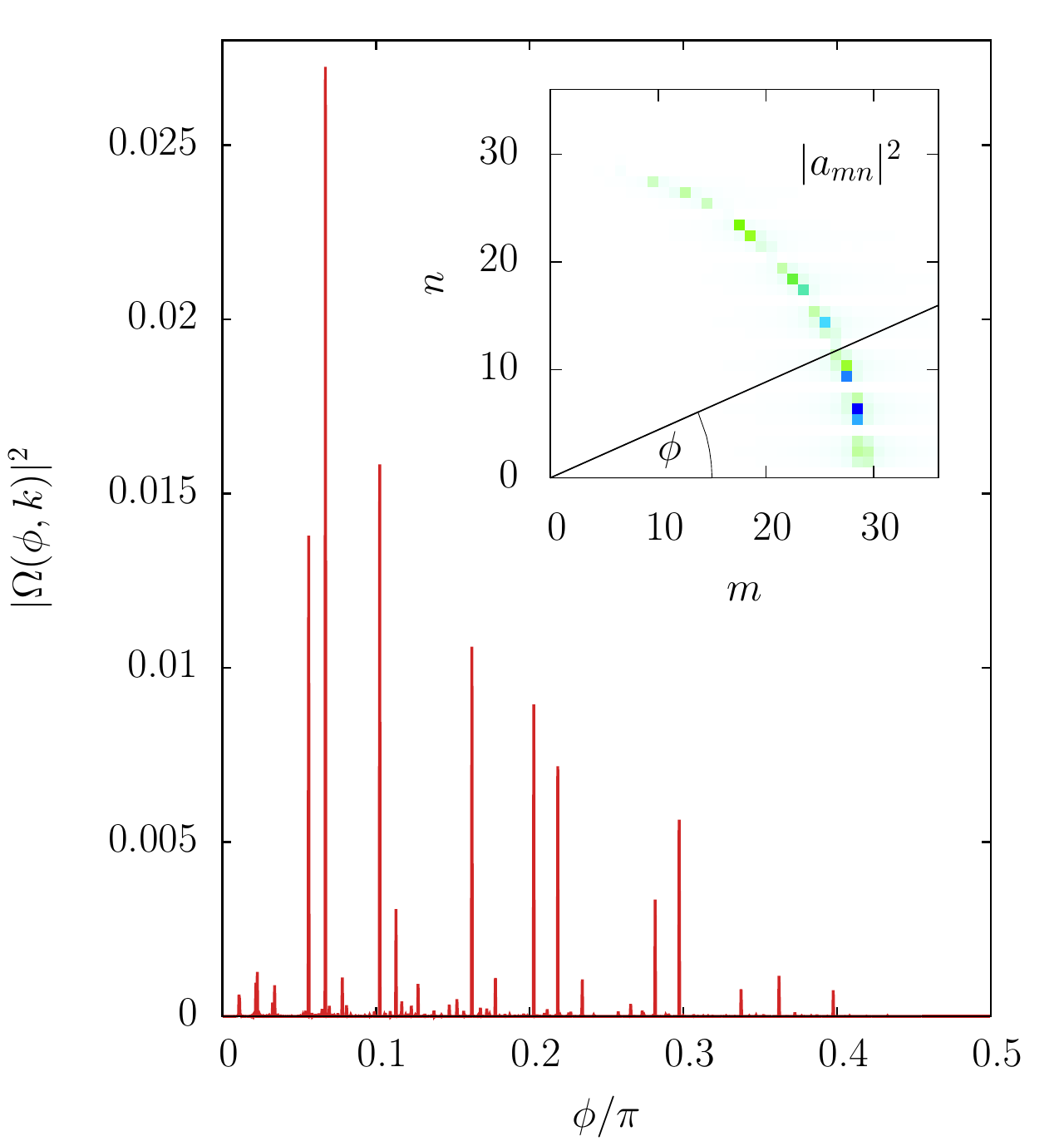}
\caption{(Color online) The absolute square of the spectral angular amplitude $\Omega (\phi,k)$ (Eq.~\ref{eq:Omega}) at fixed wavenumber $k=1.7835$ for $\Psi$ [Fig.~\ref{fig:wavenonreso} (a)]. The inset shows the expansion coefficients $a_{nm}(k)$ and the definition of the angle $\phi$.}
\label{fig:coeffs}
\end{figure}

We present in the following three prototypical cases of scattering wavefunctions: one in the smooth off-resonant part of the spectrum for incoming mode $n=1$ (Fig.~\ref{fig:wavenonreso}), one near a sharp resonance (Fig.~\ref{fig:wavereso}) and one for incoming mode $n=2$ (see Fig.~\ref{fig:wave2mode} below). Obviously, in the non-resonant case we find near perfect agreement between $\Psi$, $\Psi^{\rm T}$, and $\Psi^{\rm PSCA}$ while in the resonant case $\Psi^{\rm PSCA}$ strongly differs from $\Psi$ but is in close agreement with $\Psi^{\rm T}$, as anticipated. In resonant scattering a long-lived quasi-bound state is excited such that contributions from very long path lengths become essential while off-resonant wavefunctions typically show signatures associated with the dynamics of short paths.\\
The computational effort to calculate $\Psi^{\rm PSCA}$ is much larger than for the $S$ matrix elements since the path sum has to be performed for each grid point of a square lattice with small spacings $\delta x$. In order to generate high-resolution images of $\Psi^{\rm PSCA}$ with modest computational effort we employ the following trick: we expand $\Psi^{\rm PSCA}$ in terms of the analytically known wavefunctions of the closed billiard
\begin{align}
\langle \mathbf r|mn\rangle=\frac{2}{D}\sin{(K_mx)\sin{(K_ny)}},
\end{align}
where $K_{n}=\frac{\pi}{D}n$. With the help of the projection amplitude
\begin{align}\label{eq:amn}
a_{mn}^{\rm PSCA}=\langle mn|\Psi^{\rm PSCA}\rangle
\end{align}
evaluated on the grid, we can evaluate $\Psi^{\rm PSCA}$ in the interior as
\begin{align}\label{eq:psi_projection_modes}
\Psi^{\rm PSCA}(\mathbf r,k)=\frac{2}{D}\sum_{mn}a_{mn}^{\rm PSCA}(k)\sin{(K_mx)}\sin{(K_ny)}.
\end{align}
Due to energy conservation the expansion coefficients $a_{mn}^{\rm PSCA}(k)$ are non-zero only near the circle $K_m^2+K_n^2=k^2$ (see, e.g., the inset of Fig.~\ref{fig:coeffs}). In the limit of a bound state of a closed billiard the amplitudes take the form of a delta function $a_{mn}(k)=\delta(k-\sqrt{K_n^2+K_m^2})$. We can thus use a coarse grid $\delta x\lesssim \pi/k$ such that the maximal resolved wavenumber $\Delta K=\pi/\delta x$ is slightly larger then the wavenumber $k$ of the scattering state. $a_{mn}(k)$ calculated from this coarse grid contains the complete information on the wavefunction inside the cavity. We note that the eigenfunctions of the closed billiard do not form a complete basis for the open billiard because the lead openings are replaced by hard walls. Therefore, the Fourier expansion (Eq.~\ref{eq:amn}) and its inverse are not strictly unitary.\\
Since the expansion coefficients $a_{mn}$ are non-zero only near the circle $k^2=K_m^2+K_n^2$, we can construct a spectral angular amplitude of the scattering state with wavenumber $k$ as
\begin{align}\label{eq:Omega}
\Omega(\phi,k)=\sum_{n,m}\delta(n,\tan(\phi)m) a_{mn}(k).
\end{align}
The expression $\delta(n,\tan(\phi)m)$ is unity for $\tan(\phi)=n/m$ and zero elsewhere. Eq.~\ref{eq:Omega} can be applied to both the exact quantum state and to its semiclassical approximation. Fig.~\ref{fig:coeffs} shows the angular spectra of the non-resonant scattering state [Fig.~\ref{fig:wavenonreso} (a)]. As expected from the visual inspection of the wavefunction, the angular distribution shows many peaks corresponding to a large number of excited modes. Nevertheless $\Omega(\phi,k)$ is not uniformly distributed in angle $\phi$ because excitation of modes with large $\phi$ would lead, in general, to long lifetimes caused by their weak cavity-lead coupling [see Fig.~\ref{fig:angular}(a)]. For states with short dwell times such as Fig.~\ref{fig:wavenonreso} large $\phi$ contributions are suppressed. \\
\begin{figure*}
\includegraphics[width=\textwidth]{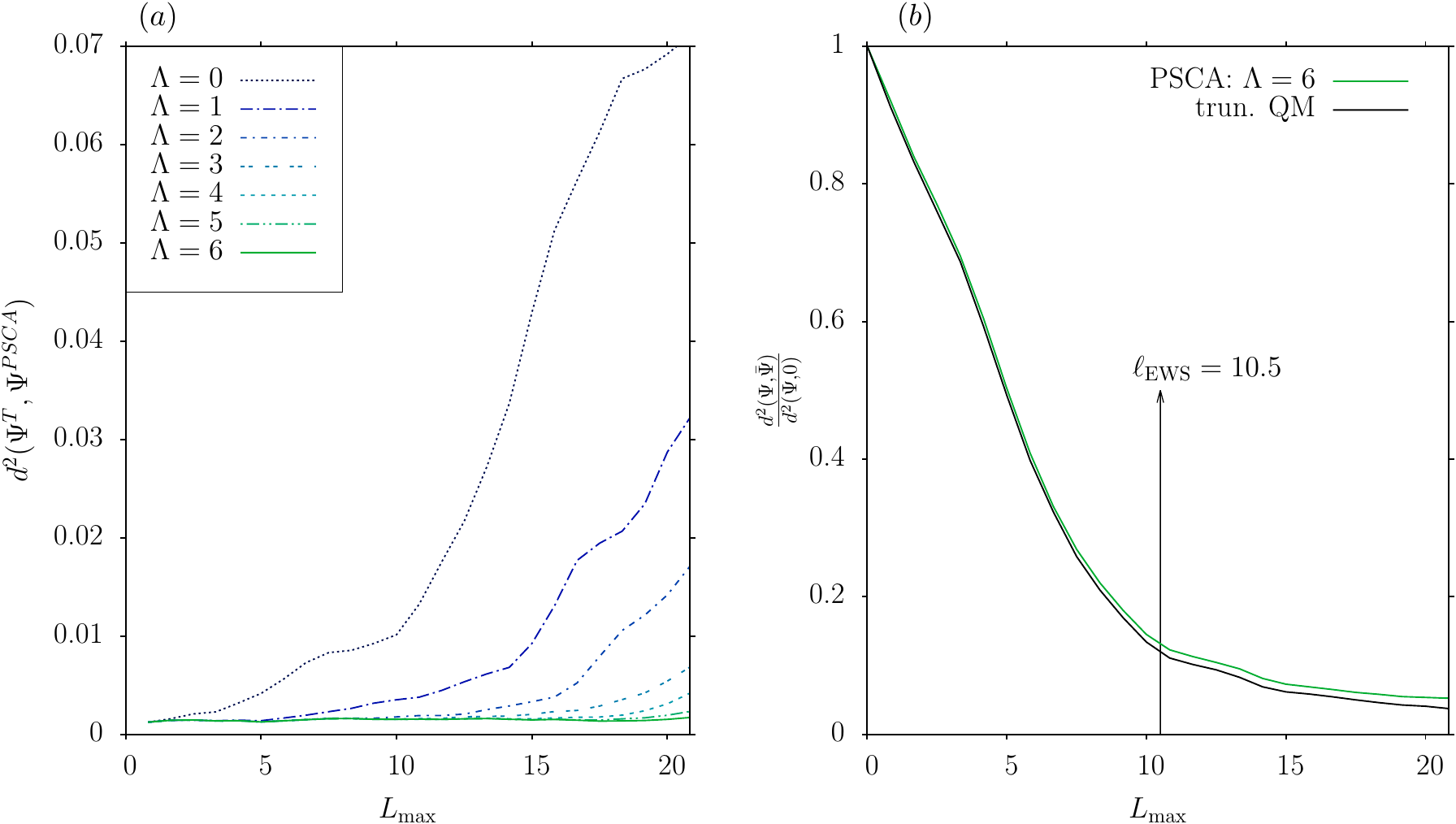}
\caption{(Color online) (a) The distance $d^2(\Psi^{PSCA},\Psi^{\rm T})$ as a function of $L_{\rm max}$ depicted for varying orders $\Lambda=1$ to $\Lambda=6$ and wavenumber $k=1.7835$ (same as in Fig.~\ref{fig:wavenonreso}). (b) The normalized distance $\frac{d^2(\Psi,\bar\Psi)}{d^2(\Psi,0)}$ of $\bar{\Psi}$ to the exact quantum wavefunction $\Psi$ $\frac{d^2(\Psi,\bar\Psi)}{d^2(\Psi,0)}$ with $\bar\Psi$ being either $\Psi^{\rm PSCA}$ (green line) or $\Psi^{\rm T}$ (black line) as a function of $L_{\rm max}$ for the same wavenumber as in (a). The mean path length within the full QM (Eq.~\ref{eq:meanlength}) is $\ell_{\rm EWS}=10.5$ and is marked by a vertical line. The PSCA is depicted for order $\Lambda=6$. $\Psi^{\rm T}$ (black line) is obtained via a discrete Fourier transform of Eq.~\ref{eq:trunc_psi} in the interval $k \in [1.725,1.875]$ with $\delta k=0.00025$.}
\label{fig:distance}
\end{figure*}
\begin{figure*}[t]
\includegraphics[width=\textwidth]{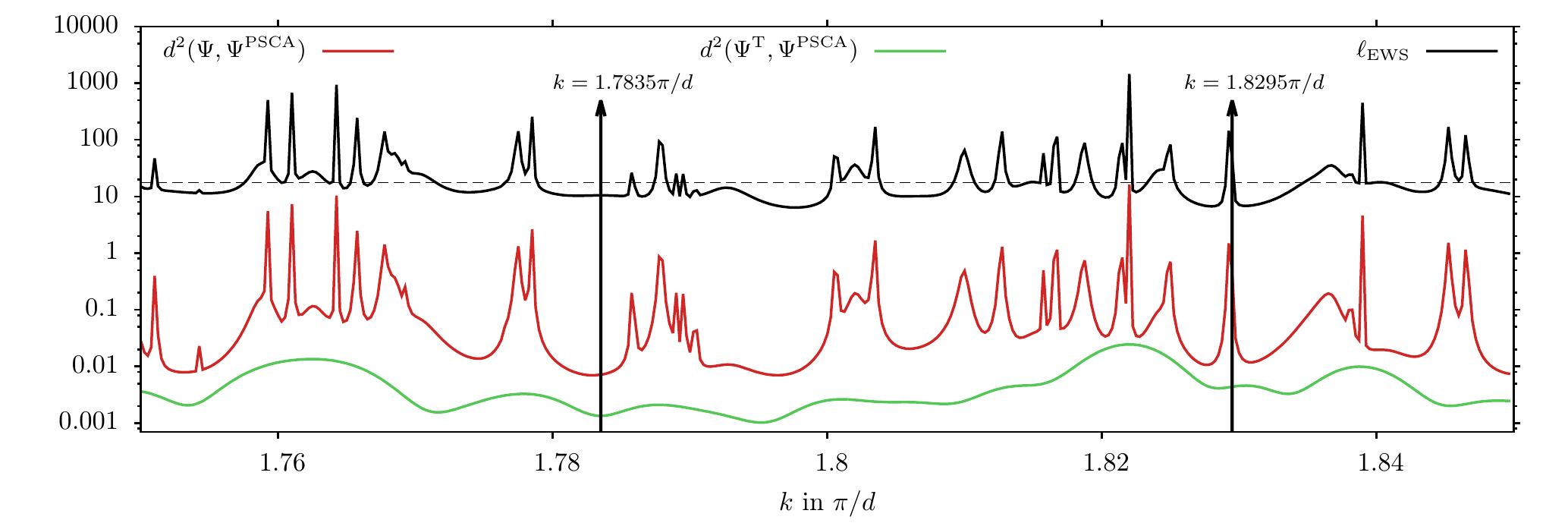}
\caption{(Color online) Distance $d^2(\Psi,\Psi^{\rm PSCA})$ between the quantum wavefunctions $\Psi$ and the PSCA $\Psi^{\rm PSCA}$ [red (dark gray) line], as well as the distance $d^2(\Psi^{\rm T},\Psi^{\rm PSCA})$ between the truncated QM and the PSCA [green (light gray)line]. The PSCA is calculated in order $\Lambda=6$ and with $L_{\rm max}=17.5$. The black solid line is the mean path length $\ell_{\rm EWS}$ (see Eq.~\ref{eq:meanlength}) and the horizontal dashed line marks a mean path length of $\ell=L_{\rm max}=17.5$.  Note the logarithmic scale of the figure. The two vertical arrows mark, respectively, the wavenumber of a non-resonant scattering state ($k=1.7835$, same as in Fig.~\ref{fig:wavenonreso}) and the wavenumber of a resonant scattering state ($k=1.8295$, Fig.~\ref{fig:wavereso}). The sharp peak at $k=1.82925$ corresponds to the resonance $m=29,n=4$ and has a mean path length of $\ell_{\rm EWS}=145$ almost degenerate to the resonance state $m=4,n=29$ at $k=1.8295$.}
\label{fig:com}
\end{figure*}
\begin{figure*}
\includegraphics[width=\textwidth]{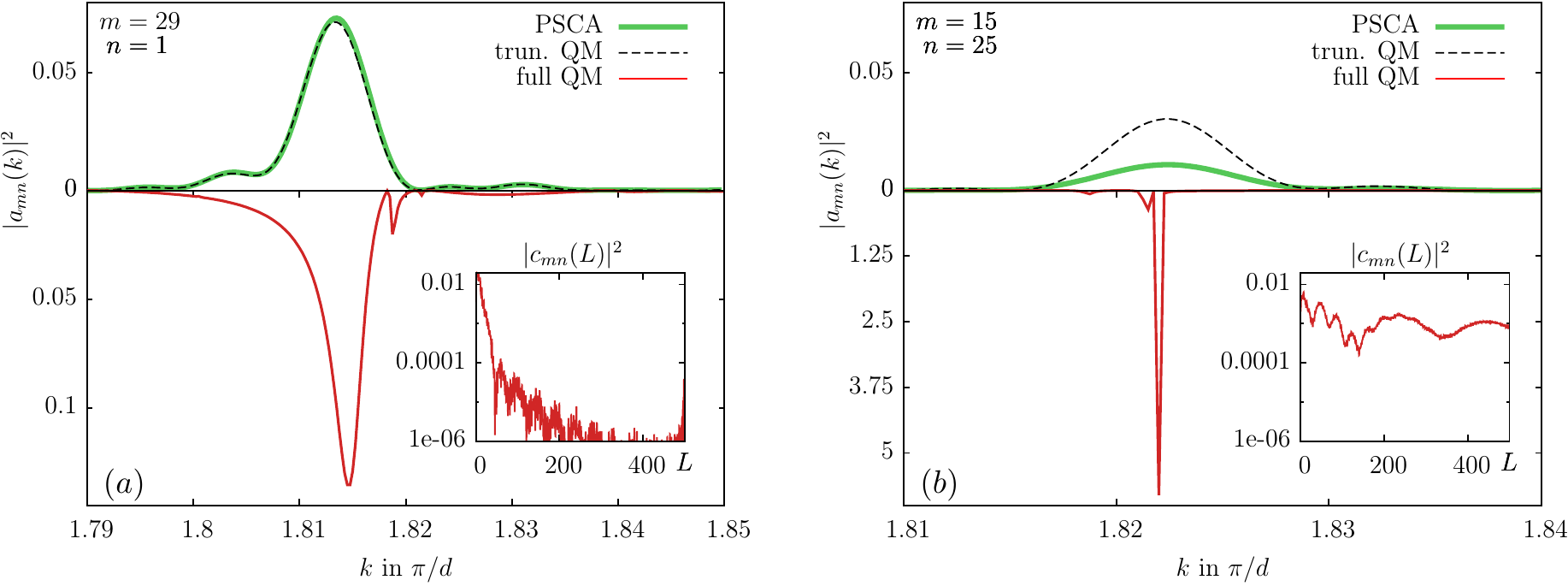}
\caption{(Color online) The absolute magnitude square of the amplitudes $|a_{mn}(k)|^2$ for the two extreme cases of (a) a broad resonance ($m=1,n=29$) and (b) a sharp resonance ($m=25,n=15$). Note the different ranges of the ordinate in (a) and (b) for the quantum result. The inset shows the length spectrum $|a_{mn}(L)|^2$ of the amplitude of $\Psi$ in a logarithmic scale. The length spectrum of the sharp resonance [inset of (b)] is not sufficiently decreased at the boundary of numerical resolution $\Delta L=500$ such that some paths with $L>L_{\rm max}$ are not filtered out of $\Psi^{\rm T}$. This causes the difference between $\Psi^{\rm T}$ (black dashed line) and $\Psi^{\rm PSCA}$ (green line) in (b). The scattering state which excites the sharp resonance corresponds to the mean path length of $\ell=1460$ at $k=1.822$ (see Fig.~\ref{fig:com}). The PSCA is calculated with $L_{\rm max}=17.5$ and order $\Lambda=6$.}
\label{fig:reso}
\end{figure*}
\begin{figure*}
\includegraphics[width=\textwidth]{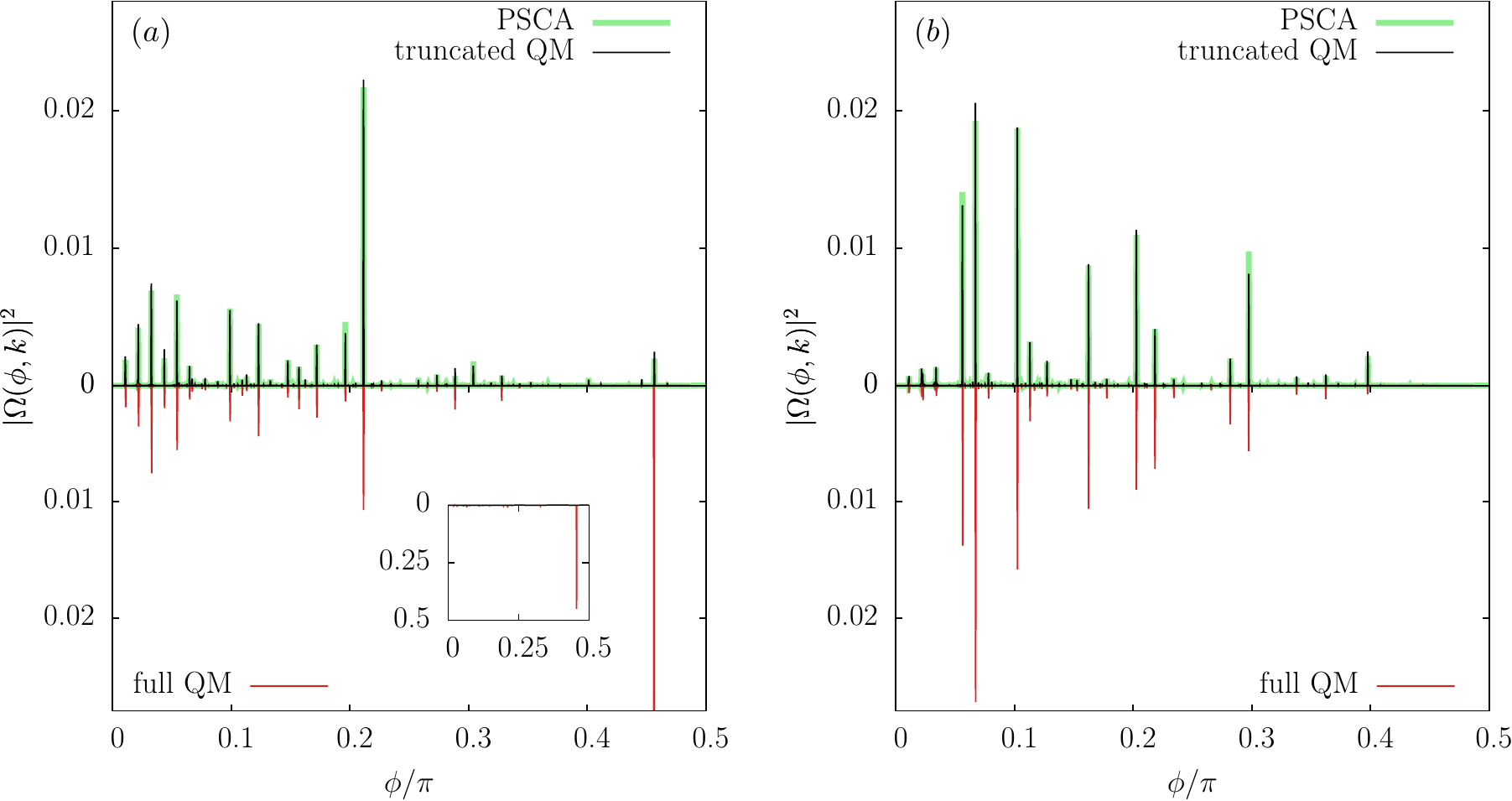}
\caption{(Color online) Comparison of $\Omega(\phi,k)$ within PSCA, truncated QM and full QM for a non-resonant ($k=1.7835$, Fig.~\ref{fig:wavenonreso}) and a resonant ($k=1.8295$, Fig.~\ref{fig:wavereso}) scattering state. The wavenumbers correspond to those labelled in Fig.~\ref{fig:com}. $L_{\rm max}=17.5$, $\Lambda=6$.}
\label{fig:angular}
\end{figure*}
\begin{figure*}
\includegraphics[width=\textwidth]{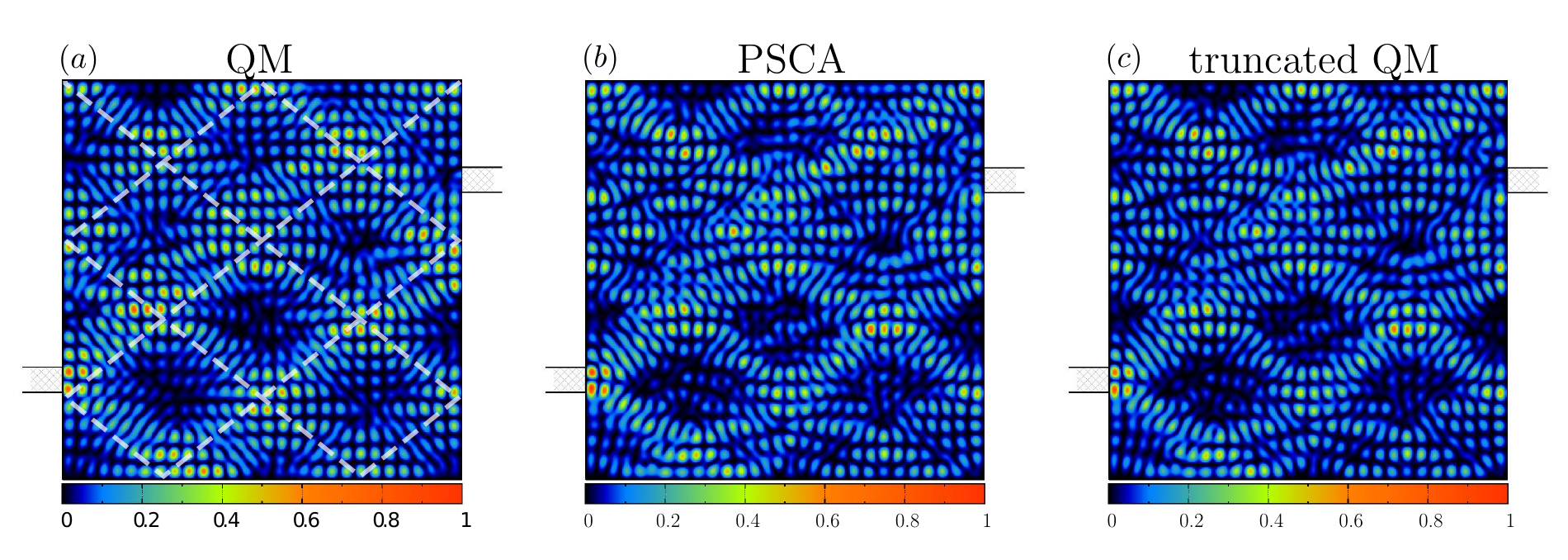}
\caption{(Color online) As Fig.~\ref{fig:wavenonreso} but for an off-resonant scattering state in incident mode $n=2$ with $k=2.25375$ and $\ell_{\rm EWS}=9.5$. Comparison of $\Psi^{\rm PSCA}$, the truncated wavefunction $\Psi^{\rm T}$, and $\Psi$. The white dashed line in (a) shows the density enhancement along the classical periodic orbit.}
\label{fig:wave2mode}
\end{figure*}
To quantify the agreement between the quantum and semiclassical wavefunctions we use the distance metric in Hilbert space
\begin{align}\label{eq:d2}
d^2(\Psi_1,\Psi_2)=\frac{1}{D^2}\iint\limits_{\rm cavity} \text{d}x\, \text{d}y \; \vert \Psi_2 (\mathbf{r},k)-\Psi_1 (\mathbf{r},k)\vert^{2}.
\end{align}
The convergence of the semiclassical wavefunction $\Psi^{\rm PSCA}$ as a function of $\Lambda$ and $L_{\rm max}$ can be conveniently studied by employing $d^{2}( \Psi^{\rm T},\Psi^{\rm PSCA})$. We observe that at fixed $L_{\rm max}$ and increasing $\Lambda$, $\Psi^{\rm PSCA}$ converges monotonically to $\Psi^{\rm T}$ which is shown in Fig.~\ref{fig:distance} (a) for the scattering state at $k=1.7835$. The semiclassical wavefunction $\Psi^{\rm PSCA}$ has converged to $\Psi^{T}$ within $d^{2}( \Psi^{\rm T},\Psi^{\rm PSCA})=0.0015$ for $\Lambda=6$ at $L_{\rm max}=20$. \\
At fixed $\Lambda$ and increasing $L_{\rm max}$ the semiclassical wavefunction $\Psi^{\rm PSCA}$ starts to diverge from $\Psi^{\rm T}$ with the onset of divergence shifted to larger $L_{\rm max}$ as $\Lambda$ increases [Fig.~\ref{fig:distance}(a)]. This is a consequence of the increasing lack of pseudopaths required for the complete path sum at length $L_{\rm max}$. For large $L_{\rm max}$ an exponentially increasing number of pseudopaths of the same length become accessible that consist of segments of shorter classical paths joined by an increasing number of diffractive scatterings (see Eq.~\ref{eq:GPSCA_part}). Some of these high-order pseudopaths are missing for the proper interference with the paths already included in the PSCA of low order $\Lambda$. Therefore, convergence to the quantum wavefunction requires a correlated limit of both $L_{\rm max}\rightarrow\infty$ and $\Lambda\rightarrow\infty$. It should be noted that with increasing order $\Lambda$, the PSCA becomes more and more sensitive to the accuracy of $v(\theta',\theta,k)$ (the diffractive amplitude for the internal scattering at the cavity-lead junction) since any error in $v(\theta',\theta,k)$ is exponentiated to the power $\Lambda$.\\
While the dependence of $d^{2}( \Psi^{\rm T},\Psi^{\rm PSCA})$ on the order $\Lambda$ of the PSCA allows to estimate the significance of diffractive contributions to the path sum up to a given length $L_{\rm max}$, the distance $d^{2}(\Psi,\Psi^{\rm T})$ measures the total contribution of paths with length beyond $L_{\rm max}$ [Fig.~\ref{fig:distance}(b)]. The convergence of $\Psi^{\rm T}$ to the exact scattering state as a function of $L_{\rm max}$ is controlled by the mean path length $\ell_{\rm EWS}$ of the scattering state. In general, for $L_{\rm max} \gtrsim \ell_{\rm EWS}$ the truncated wavefunction $\Psi^{\rm T}$, and thereby $\Psi^{\rm PSCA}$ for sufficiently high $\Lambda$ converge to the full scattering state. \\
The PSCA fails to reproduce the quantum scattering state near a sharp resonance (Fig.~\ref{fig:wavereso}) because long paths well beyond $L_{\rm max}$ contribute to this long-lived quasi-bound state. It is therefore instructive to directly compare the distance functions $d^{2}(\Psi,\Psi^{\rm PSCA})$ with the EWS length $\ell_{\rm EWS}$ as a function of $k$ (Fig.~\ref{fig:com}). Indeed, the distance of the wavefunction in Hilbert space strongly correlates with the mean path length $\ell_{\rm EWS}$ and allows to predict the accuracy of $\Psi^{\rm PSCA}$ for a given $L_{\rm max}$ when $\ell_{\rm EWS}$ is known. The two prototypical cases shown above are marked in Fig.~\ref{fig:com}. It is also instructive to measure the distance between $\Psi^{\rm PSCA}$ and $\Psi^{\rm T}$ (Fig.~\ref{fig:com}). It is uniformly small ($<0.05$) over the entire $k$ interval and slowly varying. It is worth noting that the residual fluctuations in $d^{2}(\Psi^{\rm T},\Psi^{\rm PSCA})$ are not due to shortcomings of the PSCA but rather due to the truncation process in $\Psi^{\rm T}$ (see Eq.~\ref{eq:trunc_psi}). Near resonances $\tilde\Psi(\mathbf r, L)$ contains non-negligible contributions up to very long paths. Consequently, when $\tilde\Psi(\mathbf r, L)$ is still non-negligible for $L$ beyond the Fourier resolution limit $L>\Delta L=2\pi/\delta k$, back folding causes truncation errors. The latter are responsible for the increase of $d^{2}(\Psi^{\rm T},\Psi^{\rm PSCA})$ in the region of high density of sharp resonances (e.g.~near $k=1.765$ and $k=1.822$).\\
The contribution from long paths and the possible contamination by back folding can be conveniently monitored by the projection amplitudes $a_{mn}(k)$ onto bound states of the closed billiard (Eq.~\ref{eq:amn}) or onto quasi-bound states of the open billiard. We present $|a_{mn}(k)|^2$ and its path length spectrum $|\tilde a_{mn}(L)|^2$, where 
\begin{align}
\tilde a_{mn}(L)=\int \text{d}k\; a_{mn}(k)e^{-ikL}
\end{align}
for two extreme cases in Fig.~\ref{fig:reso}, the broad resonances ($m=29$, $n=1$) and the sharp resonances ($m=15$, $n=25$). For the broad resonance, PSCA and the truncated quantum state agree perfectly since the path length spectrum of the amplitude has already decayed by over four orders of magnitude near $L=\Delta L= 500$. For the sharp resonance with $\ell_{\rm EWS}=1460$, the path length spectral intensity at $L\approx500$ is still $\approx 10^{-1}$ of its value for small $L$ and, consequently, truncation leads to large discrepancies between $\Psi^{\rm PSCA}$ and $\Psi^{\rm T}$. Naturally both the PSCA and the quantum state truncated at $L_{\rm max}=17.50$ fail to describe the exact quantum scattering state for the sharp resonance.\\
The proximity of the scattering wavefunction to a bound state of the closed billiard can also be monitored by the angular spectrum (Eq.~\ref{eq:Omega}). The sharp resonance depicted in Fig.~\ref{fig:wavereso} is very close to ($m=4$, $n=29$). Consequently, a single peak strongly determines the angular spectrum at $\tan\phi=n/m$ overshadowing all other contributions [see Fig.~\ref{fig:angular} (a), note the different scale in the inset]. For the broad resonance in Fig.~\ref{fig:wavenonreso} the angular spectrum contains many components of comparable magnitude  [see Fig.~\ref{fig:angular} (b)]. In this case, $\Omega^{\rm PSCA}$, $\Omega^{\rm T}$, and $\Omega$ closely agree with each other. \\
While the previous numerical examples feature low wavenumbers $k<2$ with only one open channel we have checked on the convergence of $\Psi^{\rm PSCA}$ for higher wavenumbers up to $k=5$ as well. We find, in general, the same very good agreement between the PSCA and the full quantum mechanics in the regime $L_{\rm max}>\ell_{\rm EWS}$ and between the PSCA and the truncated quantum mechanics in the regime $L_{\rm max}<\ell_{\rm EWS}$.\\
An example for the wavefunction convergence for the second mode with $k=2.25375$ and $\ell_{\rm EWS}=9.5$ is shown in Fig.~\ref{fig:wave2mode}. One remarkable feature of this wavefunction different from the previous cases is the intensity enhancement along a classical periodic orbit (see Fig.~\ref{fig:wave2mode}). Both $\Psi^{\rm PSCA}$ and $\Psi^{\rm T}$ show clear traces of this orbit. For the full quantum state $\Psi$ the occurrence of this structure is caused by the excitation with nearly equal amplitude of two almost degenerate eigenstates of the closed system $\vert m_i,n_i\rangle$ with $m_1=26, n_1=25$ and $m_2=30, n_2=20$. Superposition of these two eigenstates lead to an envelope that follows the track of the periodic orbit similar to the formation of a beat in acoustics \cite{CheHuaLan02}. It is now instructive to analyse their appearance within the PSCA. We first note that the angular spectrum of $\Psi$ (analogous to Fig.~\ref{fig:angular}) features two dominant peaks with equal height near the angles $\phi_i=\tan^{-1}(n_i/m_i)$ ($i=1,2$) determined by the quantum numbers of the two resonant states. Moreover the angular distribution for diffractive coupling from the point contact into the billiard in second mode $c_2(\theta,k)$ (Eq.~\ref{eq:psi_psca}) features a high amplitude near these angles, $\theta^e=\phi_i$. This differs from the coupling in first mode for which large angles would be suppressed. The build-up of the beat pattern results now from the constructive interference between the rays emitted from the contact into the billiard with entrance angle close to $\phi_i$ and its replica propagating into the same direction after 4 consecutive reflections at the billiard wall. The path length difference between two such trajectory inside the billiard is given by 
\begin{equation}
\Delta L=2 D \frac{1+\tan \phi_i}{\sqrt{1+\tan^2 \phi_i}}=2 D \frac{m_i+n_i}{\sqrt{m_i^2+n_i^2}}.
\label{eq:deltaL}
\end{equation}
The necessary condition for constructive interference 
\begin{equation}
\Delta S=k \Delta L=2\pi j \;\;\; (j \in \mathbb{N})
\end{equation}
is now approximately satisfied for both $\phi_1$ ($k \Delta L=100.01\pi$) and $\phi_2$ ($k \Delta L=101.97\pi$) at $k=2.25375$. In addition the condition for constructive interference between the rays with entrance angle $\theta^e=\phi_i$ and $\theta^e=-\phi_i$ leads to $\sin(h k_y)=0$, which gives a primitive estimate for the strength of the coupling between eigenstate and incoming second mode (for definition of $h$ see Fig.~\ref{fig:geo}). Both requirements are approximately met for $(m_i,n_i)_{i=1,2}$ resulting in the simultaneous excitation of the corresponding eigenstates and, in turn, in the beat pattern and in the density enhancement near the periodic orbit. The strong coupling between the incident mode $n$ and the two eigenstates is facilitated by the spacial proximity, to within a de-Broglie wavelength, of one of its impact points at the wall to the location of the entrance lead. \\
The excitation amplitude of the eigenstate also depends on the position of the exit lead. Taking into account higher-order diffractive corrections incorporates the coupling to the exit lead into the semiclassical description.
Note that the present mechanism for the formation of this periodic orbit is different from the appearance of scars in wavefunctions of open chaotic billiards for low $k$ \cite{AkiFerBir97} which is based on constructive interference for consecutive retracing of a single isolated periodic orbit. It is also different from the appearance of path bundles for large $k$ where bundles of short classical scattering trajectories emanating from the entrance opening whose width is determined by the lead width appear in scattering states \cite{IshKea04,RotAmbLib11}.\\
The comparison between the truncated and the full QM wavefunction (Fig.~\ref{fig:wave2mode})shows that there are non-vanishing contributions from paths with length $L>L_{\rm max}$ indicating the presence of resonances in the immediate vicinity. Investigation of $d^2(\Psi,\Psi^{\text{T}})$ as function of the truncation length $L_{\rm max}$ shows rapid decay for small $L_{\rm max}$ and a slow, plateau-like, decay for $L_{\rm max}>11$ due to the residual influence of nearby resonances which causes a small $\ell_{\rm EWS}$ but still non-vanishing distance $d^2(\Psi,\Psi^{\text{T}})$ at $L_{\rm max}=17.5>\ell_{\rm EWS}$. 

\section{Summary and Outlook}\label{sec:sum}
We have presented the construction of the semiclassical constant energy propagator and of scattering states employing the pseudopath semiclassical approximation (PSCA). The convergence  of the PSCA to the quantum limit is controlled by the maximum path length $L_{\rm max}$ and the maximum order $\Lambda$ of non-classical diffractive scatterings included. For the open rectangular billiard we find unprecedented quantitative agreement between $\Psi^{\rm PSCA}$ and the full quantum scattering state $\Psi$ when the mean path length $\ell_{\rm EWS}=k\tau_{\rm EWS}$ determined by the Eisenbud-Wigner-Smith (EWS) time delay $\tau_{\rm EWS}$ is covered by the PSCA, $\ell_{\rm EWS}\lesssim L_{\rm max}$. Thus, the pseudopaths resulting from sequences of classical paths joined by non-classical diffractive scatterings at the lead openings provide the necessary complements to the classical paths for completion of the Feynman path sum for quantum propagation. To our knowledge, this is the first protocol for constructing semiclassical scattering wavefunctions whose convergence to its quantum counterpart can be quantitatively controlled. Even when the mean path length of the exact scattering state $\ell_{\rm EWS}$ exceeds the maximum path length included in the numerical implementation of the PSCA as it happens for energies (or wavenumbers $k$) near long-lived resonances, we find near-perfect agreement between $\Psi^{\text{PSCA}}$ and the corresponding quantum wavefunction $\Psi^{\text{T}}$ with the path length spectrum truncated at the same $L_{\rm max}$ as the PSCA.
We conclude by pointing to a few future applications and extensions: due to the mathematical equivalence between the Schr\"odinger and the Helmholtz equations the experimental measurement of such wavefunctions can be conducted in open microwave billiards using movable antennas \cite{KimBarSto02} or in micro cavity lasers \cite{CheHuaLan02}. The direct measurement of wavefunctions in quantum dots is still a major challenge but some progress has been made using scanning tunneling microscopy in graphene quantum dots \cite{SubLibLi12}. The truncation of long paths beyond $L_{\rm max}$, introduced here to control the sum over an exponentially proliferating set of pseudopaths, does have, in fact, experimental analogues and applications: finite energy resolution of the detection and/or excitation processes leading to smearing out of the sharp resonances is equivalent to suppressing long paths in the expectation value $\langle |\Psi(\mathbf r)|^2\rangle_E$. Moreover, the finite phase coherence length $\ell_{\phi}$ present in decohering systems precludes the appearance of long-lived resonances and causes long paths to contribute only incoherently \cite{MarWesHop93,BarRotLib10,BloZoz00}. The latter can easily be incorporated within the PSCA by an exponential damping $~e^{-L/\ell_{\phi}}$ of long path contributions \cite{IshBur95}. Damping of long paths naturally occurs in billiards with leaky boundary conditions where tunneling through the billiard walls is present.  In the experiment such systems can be realized as microwave billiards with dielectric boundaries. The PSCA can be used to describe such systems by incorporating an additional reflection amplitude for each bounce off the billiard walls accounting for the tunneling probability. For such billiards, violation of flux conservation (unitarity) as well as a small mean path length, both naturally incorporated within the PSCA, are key features of the scattering system. The appearance of total reflection for small incident angles causes the reflection amplitude to become a pure phase factor, i.e.~its modulus is unity. In the ray picture the associated phase shift can be interpreted by a spatial shift known as the Goos-H\"anchen shift \cite{AshBir86,Jac98}. The inclusion of this effect can lead to a considerable change in the dynamics of billiards with penetrable walls \cite{ChoLeaCha94,FosCooJen07,UntWieHen08,SchHen06}. Due to the possibility of controlling individual path contributions, the PSCA has the potential to develop into an accurate method for calculating scattering states in open billiards where both diffraction at lead edges and dielectric boundary conditions are present.
\section*{Acknowledgments}
We thank Philipp Ambichl, Stefan Rotter, and Ludger Wirtz for helpful discussions. This work was supported by the FWF doctoral program ``CoQuS". Calculations have been performed on the Vienna Scientific Cluster 1.
\appendix
\section{Diffraction amplitudes}\label{sec:app}
We reproduce here the diffraction amplitudes $v(\theta',\theta,k)$ and $c_m(\theta,k)$ within the GTD-UTD. The derivation can be found in Ref.~\onlinecite{IvaWirRotBur10}.\\
The diffraction amplitude $v^{\rm GTD}(\theta',\theta,k)$ for backscattering into the cavity within the GTD \cite{Kel62} is given by 
\begin{eqnarray}
v^{\rm{GTD}}(\theta',\theta,k,d)&=& \frac{1}{2}D_L(\theta',\theta)e^{-ik\frac{d}{2}(\sin\theta'+\sin\theta)}\nonumber \\ &+&
\frac{1}{2}D_R(\theta',\theta)e^{+ik\frac{d}{2}(\sin\theta'+\sin\theta)}. \nonumber \\
\label{eq:vgtd}
\end{eqnarray}
with the scattering coefficients at the left and right wedge
\begin{eqnarray}
D_L(\theta',\theta) & = & D(\pi/2-\theta',\pi/2-\theta) \nonumber \\
D_R(\theta',\theta) & = & D(\pi/2+\theta',\pi/2+\theta)
\label{eq:gtdampl}
\end{eqnarray}
and
\begin{eqnarray}
D(\phi',\phi)&=& -2\frac{\sin{\pi/N}}{N} \Bigg[\frac{1}{\cos{\frac{\pi}{N}}-\cos{\frac{\phi'-\phi}{N}}}\nonumber \\ &-&
\frac{1}{\cos{\frac{\pi}{N}}-\cos{\frac{\phi'+\phi}{N}}}\Bigg]
\label{eq:D}
\end{eqnarray}
with $N=3/2$ the exterior angle (in units of $\pi$) of a perpendicular wedge.
The angles $\theta$ and $\theta'$ are depicted in Fig.~\ref{fig:geo}.
We use the diffraction coefficient within the UTD \cite{KouPat74} to take into account multiple scatterings between the two wedges of the cavity-lead junction:
\begin{eqnarray}
D^{\rm{UTD}}(\phi',\phi,r',r,k)=-\frac{e^{i\frac{\pi}{4}}}{N}&\times& \nonumber \\
\sum_{\sigma,\eta =\pm 1} \sigma\cot{\left(\frac{\pi+\eta(\phi'-\sigma\phi)}{2N}\right)} &\times&\nonumber \\
F\left(k\frac{rr'}{r+r'}a_\eta(\phi'-\sigma\phi)\right),
\label{eq:dutd}
\end{eqnarray}
where $a_{\pm}(\beta)=2\cos^2{\Big(\frac{2\pi Nn^{\pm}-\beta}{2}\Big)}$ and 
$n^{\pm}$ is the 
integer which most closely satisfies $2\pi Nn^{\pm}-\beta=\pm \pi$.
The function $F$ is defined as a generalized
Fresnel integral:
\begin{equation}
F(x)=-2i\sqrt{x}e^{-ix}\int_{\sqrt{x}}^{\infty} d\tau e^{i\tau^2}.
\end{equation}
With the notation
\begin{eqnarray}
U_L(\theta',\theta,r,k) & = &D^{\rm{UTD}}(\pi/2-\theta',\pi/2-\theta,r' \rightarrow \infty,r,k) \nonumber \\
U_R(\theta',\theta,r,k) & = &D^{\rm{UTD}}(\pi/2+\theta',\pi/2+\theta,r' \rightarrow \infty,r,k)\nonumber \\
\label{eq:Utd}
\end{eqnarray}
we obtain for the diffraction amplitude $v(\theta',\theta,k)$ within the GTD-UTD:
\begin{widetext}
\begin{eqnarray} 
v(\theta',\theta,k,d)  =  v^{\rm{GTD}}(\theta',\theta,k,d)+ \nonumber \\
\frac{1}{4} \sum_{\rm{odd}:\ j=1}^{j_{\rm max}}  U_L(\theta',-\pi/2,jd,k)g_j(k)e^{i\Phi^{-+}}D_R(+\pi/2,\theta)  +   U_R(\theta',+\pi/2,jd,k)g_j(k)e^{i\Phi^{+-}}D_L(-\pi/2,\theta)+\nonumber \\
\frac{1}{4} \sum_{\rm{even}:\ j=1}^{j_{\rm max}} U_R(\theta',+\pi/2,jd,k)g_j(k)e^{i\Phi^{++}}D_R(+\pi/2,\theta)  + 
U_L(\theta',-\pi/2,jd,k)g_j(k)e^{i\Phi^{--}}D_L(-\pi/2,\theta) \nonumber \\
\label{eq:vgtdutd}
\end{eqnarray}
\end{widetext}
where 
\begin{equation}
g_j(k)=\frac{1}{\sqrt{2\pi kjd}}\frac{1}{2^{j-1}}e^{i(kjd+(j-1)\pi)}
\label{gkj}
\end{equation}
and
\begin{equation}
\Phi^{\pm \pm}=k\frac{d}{2}(\pm \sin{\theta'}\pm \sin{\theta}).
\end{equation}
The sum goes over multiples of scatterings between the wedges and is cut at $j_{\rm max}$ where convergence is reached. We use $j_{\rm max}=5$.\\
The diffraction amplitude for coupling of the quantum lead to a cavity within the GTD is given by
\begin{eqnarray}
c_m^{\rm{GTD}}(\theta,k,d)&=&\frac{-ie^{\frac{im\pi}{2}}}{\sqrt{2d}k_{x,m}}\Big[\frac{1}{2}D_L(\theta,\theta_m)e^{i\frac{m\pi}{2}}e^{-ik\frac{d}2\sin\theta}\nonumber \\ &-&
\frac{1}{2}D_R(\theta,\theta_m)e^{-i\frac{m\pi}{2}}e^{ik\frac{d}2\sin\theta}\Big].
\label{eq:dngtd}
\end{eqnarray} 
As before we use the notation
\begin{eqnarray}
D_L(\theta,\theta_m) & = & D(\frac{\pi}{2}-\theta,\frac{3\pi}{2}-\theta_m), 
\nonumber \\
D_R(\theta,\theta_m) & = & D(\frac{\pi}{2}+\theta,\frac{3\pi}{2}-\theta_m).
\end{eqnarray}
for the left and right wedge. The angle $\theta_m$ is determined by the open lead mode and is given by $\theta_m=\arcsin(m\pi/dk)$. Using the UTD Eq.~\ref{eq:Utd} for the out-coupling into the cavity we obtain:
\begin{widetext}
\begin{eqnarray} 
c_m^{\rm{GTD-UTD}}(\theta,k,d)=c_m^{\rm{GTD}}(\theta,k,d) -i \frac{e^{\frac{im\pi}{2}}}{\sqrt{2d}k_{x,m}}&\times& \nonumber \\
\frac{1}{4}\Bigg[\sum_{\rm{odd}:j=1}^{j_{\rm max}}U_R(\theta,+\pi/2,jd,k)g_j(k)e^{i\phi^{++}}D_L(-\pi/2,\theta_m)-
U_L(\theta,-\pi/2,jd,k)g_j(k)e^{i\Phi^{--}}D_R(+\pi/2,\theta_m)&-& \nonumber \\
\sum_{\rm {even}:j=1}^{j_{\rm max}}U_R(\theta,+\pi/2,jd,k)g_j(k)e^{i\Phi^{+-}}D_R(+\pi/2,\theta_m)+
U_L(\theta,-\pi/2,jd,k)g_j(k)e^{i\Phi^{-+}}D_L(-\pi/2,\theta_m)\Bigg], \nonumber \\
\end{eqnarray}
\end{widetext}
where 
\begin{eqnarray}
\Phi^{\pm \pm}=\pm k\frac{d}{2}\sin{\theta}\pm \frac{m\pi}{2},
\end{eqnarray}
and $g_j(k)$ is given in Eq.~(\ref{gkj}).

%

\end{document}